\documentstyle[aps,floats,prl,twocolumn]{revtex}
\begin{document} 

\title{ Generalized Variable Range Hopping Near Two-Dimensional Metal-Insulator
        Transitions }
\author{Wenjun Zheng and Yue Yu}
\address{Institute of Theoretical Physics, Chinese Academy of Sciences, Beijing
       100080, P.R.China}
\date{\today}       
\maketitle 

\begin{abstract}
       In an attempt to understand quantitatively the remarkable discoveries of metal-insulator
transitions in two-dimensional systems, we generalize Mott's variable range hopping
theory to the situation with strong Coulomb interaction. In our formulation,
the Gaussian form is adopted into the expression of the hopping probability,
and the effect of Coulomb gap is also considered. After taking account of
the newly proposed scaling consideration, we produce the dynamical and localization length
exponents, which are consistent with the experiments. We then clarify the physical content of our formulation and explain the universality of the localization 
length
exponent suggested by a series of experiments. We also discuss the 
 general scaling function of both temperature and electrical field on the insulating side of the transition.

\end{abstract}
\vskip 0.2in
        The conventional scaling theory for disordered systems holds that without
 interactions any extent of disorder is sufficient to localize the electrons and 
 no true metallic behavior is possible at $T=0$ in two dimensions\cite{SCAL}. The inclusion
 of interactions to this picture proved to be much more difficult, and no satisfactory
 theoretical picture has emerged in spite of decades of continuous effort. However the belief that all the states
 are localized at $d=2$ has remained unchallenged.
 
       Recently, a series of surprising experiments have been conducted on 2D electron
gas and hole gas in zero magnetic field, which demonstrated convincing evidences
in support of a true metal-insulator transition(MIT) in 2D \cite{MIT}. In these experiments, The resistivity scales 
with temperature or electrical field with a single scaling parameter that 
approaches zero at a critical carrier density $n_c$. Further more the scaling 
function of resistivity near the
transition possesses a very simple exponential form, which leads to the reflection
symmetry across the transition, as emphasized in ref\cite{SYM}. That is,
\begin{equation}
\label{EQ1}
\rho(\delta n,T)=1/\rho(-\delta n,T), 
\end{equation}
where $\delta n=(n-n_c)/n_c$, $\rho$ is the resistivity measured in unites of
 its critical value $\rho_c$. This symmetry is very similar to some recent experimental 
 results in quantum Hall systems\cite{Hall}, although the mechanism may be
quite different\cite{ZYS}. These remarkable discoveries have evoked a wave
of theoretical efforts to understand them \cite{THEO}\cite{DOB}. Although it is generally agreed that
strong interactions play an important role, no specific microscopic mechanism
has been able to explain the experiments both qualitatively and quantitatively.

       In a recent Letter\cite{DOB}, Dobrosavljevic et al. gave a general scaling consideration
of the 2D MITs. In their formulation, they treated the temperature-dependence of
the conductance in the quantum critical region. After assuming the MIT occurs
at $g=g_c$ where $\beta(g_c)=0$ and introducing the scaling variable $t=\log(g/g_c)$,
they considered the linear approximation to $\beta(t)$ near $t=0$ and obtained
$$\beta(t)=\frac{dt}{d(\log L)}\approx \frac{t}{\nu}+O(t^2).   $$
By integrating from $l$ to $L$, they got
$$ t(L)=t_0(\frac{L}{l})^{1\over\nu},$$
where $t_0=\log(g_0/g_c)$ is given by the value $g_0$ of the conductance at the
microscopic scale $l$. Then after using $t_0\approx(g_0-g_c)/g_c\propto \delta n$
, where $\delta n=(n-n_c)/n_c$ and $n$ is the density, they reached
$$ g(L)=g_c\exp[A\delta n(L/l)^{1\over \nu}].  $$
Then representing the length scale $L$ by the temperature as $T\propto L^{-z}$, one
finally arrived at
\begin{equation}
 g(\delta n,T)=g_c\exp(A\delta n/T^{1\over{\nu z}}).  
\end{equation}
This expression gives the explicit exponential form of the scaling function of the conductance
with respect to temperature except for the exponents $\nu$ and $z$ to be determined,
and the reflection symmetry in eq(\ref{EQ1}) follows from it directly, which are precisely what has been
seen in the experiments. As emphasized by the authors\cite{DOB}, the above 
relation is only valid inside the critical region, or for $T>T_0\propto |\delta n|^{\nu z}$, 
which is also verified experimentally\cite{SYM}.
In what follows we will use this result as a restrictive
condition to our general derivation, which is the generalization of Mott's well-known 
variable range hopping (VRH) theory. Then we will arrive at the explicit scaling 
function of the resistivity, from which 
the exponents $\nu=1.5$ and $z=1$ can be read off easily, which are consistent with 
the experiments\cite{NUZ}. We then discuss the physical mechanism of our formulation and 
explain the universality of the exponent $\nu=1.5$\cite{NU}, which is characteristic of a general two
dimensional  systems with strong, long-range interactions( like Coulomb 
interaction). At last, we discuss the general scaling function of both temperature 
and electrical field. 
                
Before starting with our formulation, let us give an introduction to some
background that motivates the present work.      
        In an influential work\cite{MOTT}, Mott suggested his celebrated theory of VRH
which described an important conducting mechanism in a localized electron system. His
central idea is to optimize the hopping probability with respect to the hopping
distance $R$, which gave the temperature-dependence of the conductivity
$$\sigma(T)\propto \exp[-(\frac{T_0}{T})^{1\over{d+1}}],     $$
where the dimension $d=2$ here.
This result has received substantial experimental supports in the past decades
\cite{EP1}.
Later, Efros et al. made an attempt to include the effect of Coulomb interaction
into the VRH theory \cite{EFR}. They only considered the Coulomb gap correction to the
density of states and obtained
$$\sigma(T)\propto \exp[-(\frac{T_0}{T})^{1\over{2}}],   $$
This has also been verified in the experiments 
performed on dilute electron systems where Coulomb interaction dominates the 
transport properties \cite{EP2}. Although this relation can produce the right 
exponent $z=1$, it is nevertheless inconsistent with the concrete expressions 
derived from the accurately designed measurements performed recently\cite{MIT}
 and fails to explain the general localization length exponent $\nu\approx 
 1.5$ revealed by more than one experiment.

In the following, we will extend the VRH idea 
to the situation with strong Coulomb interaction. Our generalization goes beyond
the one attempted by Efros et al., and aims at a better understanding of the 
recently surging discoveries of MITs in two dimensional systems,
especially the exponent $\nu$, which seems to be a universal exponent.
 
 The hopping probability for a particle to cover a distance of $R$ is
\begin{equation}
\label{EQ2}
p(R)\propto \exp(-\alpha\frac{R^2}{\xi^2}-\frac{\Delta}{k_B T}),  
\end{equation}
where $\xi$ is the localization length on the insulator side of the MIT, which
diverges as $\xi\propto |\frac{n-n_c}{n_c}|^{-\nu}$ when approaching the MIT ($\nu$ is the
localization length exponent); $\alpha$ is a non-singular parameter,
$\Delta$ is the energy difference between the initial state and the final state
for the hopping particle. The only difference from Mott's equation is the Gaussian form 
dependence of $p(R)$ on $R$, which is vital to the final conclusion and contains a profound
physics content(see below).
If there is no interaction, we simply have 
$$\Delta\propto [R^dN(E_F)]^{-1},$$
as it is supposed in Mott's derivation, where $N(E_F)$ is the density of states
at the Fermi level $E_F$. However, for the case of strong Coulomb
interaction, the Coulomb gap will depress the density of states near the Fermi surface.
Following Efros et al.\cite{CG}, we can write
$$N(E)\propto |E-E_F|, $$
which means $N(E_F)=0$. 
Therefore $N(E_F)$ in the above equation should be substituted by the following 
integration which counts the number of states available for a R-range-hopping:

\begin{eqnarray}
N(E_F)&\rightarrow& \frac{1}{\Delta E}\int_{0}^{\Delta E}dE N(E+E_F) \\  \nonumber
&\propto& \Delta E   \\   \nonumber
&\propto& \frac{1}{R},
\end{eqnarray}
where a linear dispersion $\Delta E\propto\Delta k$ has been used, and $\Delta k$
is the momentum quanta proportional to $1/R$.

Therefore $\Delta\propto 1/R^{d-1} $. This means the effect of Coulomb interaction 
is to reduce the dimension by 1. Therefore in what follows, we will use the effective
dimension $d^\prime=d-1$.

Then we maximize eq(\ref{EQ2}) with respect to $R$, and obtain
$$ R_{max}\propto (\frac{\xi^2}{T})^{1\over {d^\prime+2}}.  $$
Therefore we finally arrive at
\begin{equation}
\label{CEN}
\sigma(n,T)\propto p(R_{max})\propto \exp\{-\frac{(|\Delta n|)^{{2d^\prime\nu}\over{d^\prime+2}}}{T^{2\over {d^\prime+2}}}\}.  $$
\end{equation}
This is the central equation of the present paper, which should be contrasted with
the relations of Mott and Efros et al.. 

Taking into account the classical scaling relation
$$ \sigma(n,T)=F(\frac{\Delta n}{T^b}),  $$
where $F(x)$ is the scaling function, $b=1/z\nu$, and $z$ is the dynamical exponent, we can easily see 
$$z=d^\prime=1 $$
for a two dimensional system. This result is well-known for a Coulomb interaction
dominated system\cite{Cou}. This relation is also verified in many recent experiments on 2D
MIT \cite{MIT}.

      However, there is more to obtain from eq(\ref{CEN}). Since the above
single parameter scaling argument has given the following restrictive form\cite{DOB}
$$ \rho(n,T)\propto \exp(-A{\frac{\delta n}{T^b}}),  $$
we have an additional condition to satisfy:
$$ {{2d^\prime\nu}\over{d^\prime+2}}=1, $$
which gives 
$$\nu={{d^\prime+2}\over {2d^\prime}}=1.5. $$
This is very close to the localization length exponent reported widely by a series
of experiments \cite{MIT}. Therefore,
\begin{equation}
\label{EQ3}
\rho(n,T)\propto \exp(\frac{|\Delta n|}{T^{2/3}}) .
\end{equation}

We would like to comment that even if we have adopted the
traditional $p\propto \exp({-R/\xi})$ instead of the Gaussian form, we can still
arrive at $z=1$. However the desirable $\nu$ can not be obtained in that way.
That is to say, the traditional VRH theory is not consistent with the scaling
argument presented in ref\cite{DOB}, under the condition that $\nu=1.5,~~z=1$.

Now let us give some comments on the physics contained in the Gaussian form
adopted in eq(\ref{EQ2}). This form is quite natural and reasonable in physics, because
of its direct connection with the eigenstates of the harmonic oscillator(HO), one of the few universal  
and analytically transparent models. Motions in a general smooth enough potential near its minimum
can be approximately described by a HO. Therefore, if there does exist such effective
potential that can equivalently describe the motion of the localized particles relevant to
the 2D MITs, the Gaussian form will be probably a most
suitable one to be employed. However, considering the success of Mott's original
relation and the strong Coulomb interaction present in the samples used for the
research of 2D MITs, we are more inclined to the viewpoint that the effective
potential felt by the hopping particles are probably relevant to the long-range
Coulomb interaction. First, the property of long range ensures that the potential experienced
by one particle is the sum of the contributions of many other particles, which
is averaged to be smooth and slowly changing. Secondly, the domination of Coulomb
interaction over the kinetic energy suppresses the dynamical fluctuations which push
the particles away from their most stable position with the lowest potential energy.
Therefore it will be a relatively good approximation to represent the effective
potential for a hopping particle by an HO  potential, with the particle's
initial position at the center. For comparison, we give further comments on the 
traditional form $\exp(-R/\xi)$. This form is specific to a square well potential,
which is widely used in models of localization. As we believe, the above form
is good for short-range interactions like a hard-core potential, or in a tight
bound model where electrons are bound by the ionic potential. However, in 
case of long-range interactions, the much smoother HO potential is better.  
This argument may explain why $\nu=1.5$ is extensively reported in
strongly interacting 2D systems, independent of the sample parameters.\cite{EXP1}
  
    Then we discuss what happens when an electrical field $E$ is turned on, and what
will be the explicit form of the scaling function with two variables $E$ and $T$
 on the insulating side of the MIT. With
the assistance of  electrical field $E$, the hopping probability can be written as
\begin{equation}
\label{EQ4}
p(R)\propto \exp(-\alpha\frac{R^2}{\xi^2}-\frac{\Delta}{k_B T}+\frac{ER}{k_B T}).  
\end{equation}  
With $z=1$ in mind, we can define two dimensionless scaling variables:
$$ e=E\xi^2,~~~~~t=T\xi.$$
Then suppose $R\sim\xi e^y t^x$ ($x, y$ are exponents to be decided), and substitute it into eq(\ref{EQ4}),
 we have
$$p(R)\propto \exp[F(e,t)], $$
where $F(e,t)$ is the general scaling function of $e$ and $t$, and is the
sum of the following three parts\cite{EXP}:
\begin{eqnarray}
F_1&\sim&-e^{2y}t^{2x},  \\  \nonumber
F_2&\sim&-e^{-y}t^{-x-1},  \\  \nonumber
F_3&\sim&e^{y+1}t^{x-1}.
\end{eqnarray}
We then maximize the eq(\ref{EQ4}) with respect to $R$, and get
$$\frac{2\alpha R}{\xi^2}\sim\frac{1}{R^2T}+\frac{E}{T}. $$
By changing into $e$ and $t$ the above equation can be written as
\begin{equation}
\label{CON}
e^{y}t^{x}\sim e^{-2y}t^{-2x-1}+e t^{-1}.
\end{equation}
Therefore the general scaling function is the sum of $F_1$,$F_2$ and $F_3$,
with the exponents $x$ and $y$ decided by the constriction equation (\ref{CON}).
This suggests that the scaling function will be of different  forms for
different regions of $e$ and $t$, and demonstrates complicated behaviors
as a result.

    We first discuss the simple case of $|e|<<1$. In order for $F(e,t)$ to
give a convergent result, we must satisfy:
$$ 2y\geq 0,~~~-y\geq 0,~~~y+1\geq 0. $$
Therefore it is necessary that $y=0$, which is substituted into eq(\ref{CON}). Then
we get $x=-2x-1$, so $x=-1/3$. We thus arrive at the scaling function for
$|e|<<1$,
$$ F(e,t)\sim -t^{-2/3}+e t^{-4/3}. $$
For $e=0$, we obtain 
$$ \rho(n,T)\propto \exp[-F(0,T)]\propto \exp(\frac{-A\delta n}{T^{2/3}}), $$
which is exactly eq(\ref{EQ3}) we get earlier.

     It is also interesting to understand the behavior of the scaling function
for $t<<1$ and $t>>1$ which will be reported elsewhere because of its complexity.     
 
     Before closing, we would like to give more comments on our use of a hopping 
theory to describe transport properties near MITs. We first point out an important 
difference of our hopping picture from Mott's original theory. In Mott's
 formulation, the relevant system is strongly localized, so the localization 
length $\xi$ is very small and the overlap between adjacent wave-packets is
also small. Therefore, in order for a hopping precess to contribute to the 
conductance, the hopping distance $R$ must be larger than $\xi$, which puts a 
constraint on $R_{max}$. This leads to the conclusion that Mott's hopping is 
only valid at low enough temperature, or $T<T_0$. However the situation near 
MITs is quite different, where $\xi$ becomes very large, and the overlap 
between adjacent wave packets is also large, so a particle hopping between 
neighboring sites can contribute to the conductance without covering a distance 
as long as $\xi$. Therefore the low-temperature-constraint on hopping mechanism 
is actually absent, and we are justified to apply it inside the critical region 
$T>T_0$. The relevance of hopping for both $T>T_0$ and $T<T_0$ is supported by inspecting the inset
of Fig.2 in ref\cite{SYM}, where no crossover behavior is detected near $T_0$ 
for the insulating phase, while the metallic phase shows obvious deviation from
the linear behavior for $T<T_0$. In fact, the use of hoppings to study critical phenomena is not
new\cite{CMR}, and its relevance to critical exponents (i.e. $z$) was already 
established by Efros et al.\cite{CG}.
   
     In conclusion, we have presented a generalization of Mott's variable range hopping
theory to the situation with strong Coulomb interaction in order to understand  quantitatively the remarkable discoveries of 
metal-insulator transitions in two-dimensional systems. In this formulation,
the Gaussian form is adopted into the expression of the hopping probability,
and the effect of Coulomb gap is also considered. After taking into account 
the newly proposed scaling consideration, we for the first time give the explicit scaling
function of the resistivity on the insulating side, and as a result determine the dynamical and localization length
exponents, which are consistent with the experiments. We then clarify the physical
meaning contained in our formulation with its relevance to the long-range Coulomb
interaction. We also discuss the general scaling function of temperature and electrical field on the insulating side of the transition.     

We acknowledge helpful discussions with Z. B. Su and V. Dobrosavljevic. This work is in part
supported by National Natural Science Foundation of China.


\begin{references}
\bibitem{NU} We derive the universality of $\nu$ from the general value of $\nu z$, because it has been widely accepted that $z\approx 1$ is a universal exponent of 2D strongly interacting systems.
\bibitem{EXP} In what follows, we use $\sim$ in equations where constant coefficients are omitted for convenience.
\end{references}
\end{document}